# Analysis of Computer Hardware Affecting Video Transmission via IEEE 1394a connection

*T. Mirzoev, Ph.D.*

*The Department of Information Technology*
*College of Information Technology, Georgia Southern University, Statesboro, GA, 30640*

## ABSTRACT

When 60 de-interlaced fields per second digital uncompressed video is streamed to a computer, some video fields are lost and not able to be stored on a computer's hard drive successfully. Additionally, this problem amplifies once multiple video sources are deployed. If it is possible to stream digital uncompressed video without dropped video fields, then a sophisticated computer analysis of the transmitted via IEEE 1394a connection video is possible. Such process is used in biomechanics when it is important to analyze athletes' performance via streaming digital uncompressed video to a computer and then analyzing it. If a loss of video fields occurs, then a quality analysis of video is not possible.

## MOBILE COMPUTER HARDWARE

The recording of video is complicated when mobility on a field is required. Seldom, desktop computers are used to collect video information out on a field due to the lack of mobility, weight and connectivity options. Laptop computers are utilized to perform the same functions with little or no impact on performance. However, the cost associated with powerful "desktop replacements" (laptops) often interferes with one's ability to perform various tasks. Therefore it is important to understand whether certain computer hardware on a laptop computer improves the transmission of video via IEEE 1394a protocol or certain hardware configurations, often expensive ones, do not improve the transmission of video. This manuscript presents the examination of the following hardware components: 1) hard disk rotational speed (RPM), 2) amount of Random Access Memory (RAM), and 3) the number of video sources that record and transmit a video file.

## SYSTEM DESIGN

In order to determine which hardware components or how many video streaming sources contribute to the loss of video fields during a DV video transmission over IEEE 1394a connection, there are several important considerations that need to be analyzed. First, a proper hardware and software setup is essential for providing DV-quality video streaming for biomechanical applications. Second, it is essential to study and examine hardware factors and video streaming sources that contribute to loss of video fields which, if the loss occurs, limits a scientist's ability to digitize and analyze captured video for biomechanical analysis. Besides the hardware design, the software package Ariel Performance Analysis System (APAS) motion analysis software is used as a tool for capturing video and performing biomechanical analyses on a computer.

In order to identify hardware components that contribute to the loss of video fields when 720x480 60 de-interlaced fields per second video is transmitted via IEEE 1394a connection to a laptop computer with installed Ariel Performance Analysis System motion analysis software, a comparative analysis of different computer hardware configurations and a number of video streaming sources are needed.

In this research study a connection between three digital video camcorders and a laptop computer is implemented. Different hardware components of the portable computer and a number of video streaming sources (DV camcorders) are the primary subjects for this discussion.

**TESTING ENVIRONMENT**

This research study's method of investigation involved several steps. In order to evaluate the contribution of the hardware elements to the loss of video fields, several test scenarios are created and represented by Table 1. Each configuration test scenario A, B, C, D, E, or F represents the steps to be taken during the experiment. The number of collections for each scenario is calculated in the following manner: a typical video transmission for 100 meter hurdles is expected to last about 12-15 seconds, so 10 seconds is used as a recording time for data collection.

Table 1

*Test configuration scenarios for Gateway 450ROG laptop.*

|  | Configuration | | | | | |
| --- | --- | --- | --- | --- | --- | --- |
|  | A | B | C | D | E | F |
| Memory Size, Mb | 1024 | 1024 | 1024 | 1024 | 1024 | 1024 |
|  | 512 | 512 | 512 | 512 | 512 | 512 |
| Hard Drive Speed, RPM | 7200 | 7200 | 7200 | 5400 | 5400 | 5400 |
| Number of Video Sources | 1 | 2 | 3 | 1 | 2 | 3 |

For example, in jumping events there are three attempts per heat with 10-12 athletes participating (30-36 trials). That creates thirty 10-second collections for each test configuration scenario A, B, C, D, E or F.

Each time a video transfer of video occurs, APAS indicates whether a loss of video fields occurs. Every transfer session results will be inputted into the population sample. Figure 1 indicates a sample screen captured from APAS after a video transfer is complete.

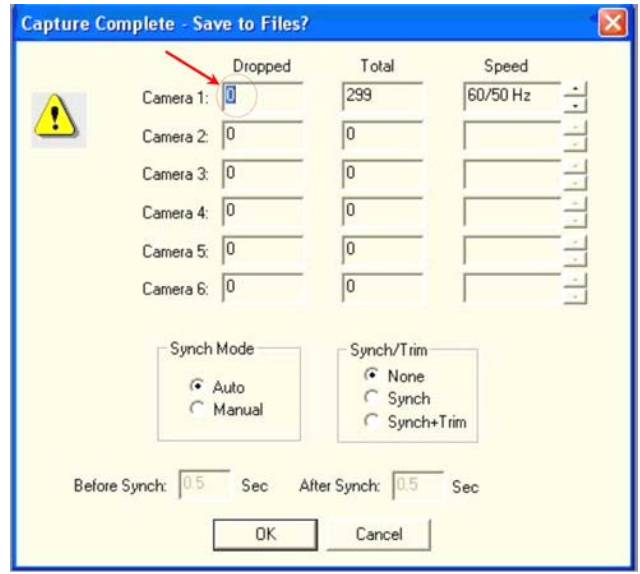

Figure 1. APAS indication of captured video fields.

The following necessary tools and applications were utilized during the data collection process:
1. APAS - Ariel Performance Analysis System motion analysis software, version 2005. This software would allow analyzing video on the laptop computer transferred through the IEEE 1394a connection from the a DV camcorder(s) (video sources),
2. Acronis True Image Home version 9.0. This application's usage is for cloning an image of a hard drive,
3. Gateway 450ROG laptop with Windows XP Professional 32-bit version,
4. Hitachi hard drives HTS541060G9A00 (5400 RPM) and HTS726060M9AT00 (7200 RPM),
5. IEEE 1394a 10m cable made by Unibrain (part# 1638),
6. COMPAQ 2 Port Firewire PCMCIA Cardbus model no. CPQFWCB (serial # UMI4103952).

**FINDINGS AND ANALYSIS OF TRANSMITTED FIELDS**

The purpose of this study was to identify hardware components that contribute to the loss of video fields when 720x480 60 de-interlaced fields per second video is transmitted via IEEE 1394a connection to a laptop computer with installed Ariel Performance Analysis System motion analysis software. A comparative analysis of different computer hardware configurations and a number of video streaming sources was performed and several outcomes of this analysis are presented in this article.

There were several variables that were considered in this study – hard disk speed (RPM), memory (RAM) and the number of video sources (VideoStreams) connected to the tested system. Based on the results of this study, the amount of RAM did not have a significant main effect of the number of video fields transmitted.

Based on the results of this study, RPM variable showed a significant main effect on the number of collected video fields. Tested configurations with 7200 RPM hard drive showed lower performance in video frames collection. Five out of six tested configurations with RPM measure equal to 5400 did not result in any dropped fields. Only one tested PC configuration with 1024 Mb of RAM and 5400 RPM resulted in dropped fields. However, 2 out of 6 tested Gateway computer configurations with 7200 RPM and 512 Mb of RAM resulted in dropped fields. Furthermore, most resourceful configurations with 1024 Mb of RAM and 7200 RPM hard disk speed resulted in collections of more than expected 300 frames. That phenomenon occurred when more than 1 video camera transmitted video to the tested Gateway laptop.

## RECOMMENDATIONS FOR FUTURE RESEARCH

The tested Gateway portable computer had an x86 Family 6 Model 13 Stepping 6 Genuine Intel ~1798 MHz processor installed that supports Stepping Technology. Perhaps, the CPU's Stepping Technology could explain why the most resourceful configurations of this study's research experiment with 1024 Mb of RAM and 7200 RPM hard disk speed resulted in collections of more than expected 300 frames. It is a possibility that the tested CPU switch to a faster clock speed when multiple video cameras were connected to the tested Gateway computer. This indicates a need to further investigate the effects of different CPU technologies' impacts on the number of collected video fields. For future research configurations, it is recommended to test the following CPU hardware configuration scenarios:
1. Test CPU with Stepping Technology support vs. regular, clock speed non-adjustable CPU,
2. Test CPU with Stepping Technology support vs. multi-core CPU According to Technology@Intel Magazine, "dual-core processor-based PCs are the next generation in PC computing performance and power Dual-core processor-based PCs are the next generation in PC computing performance and power" (Intel Corp. 2005),
3. Test the impacts of having various levels of on-chip cache memory.

Another computer component that might affect the transmission of video to APAS software via IEEE 1394a connection is storage controller type. For example, Redundant Array of Independent Disks (RAID) is still widely utilized in industry; furthermore, home computers are starting to utilize this technology of RAID due to the decreasing pricing, better availability and a range of manufacturers that offer such technologies. In this study, Gateway computer utilized Intel(R) 82801FBM Ultra ATA Storage Controller – ATA-7 with 100Mb/s throughput. However, more resourceful systems with multiple hard drives that are configured in RAID could have a different main effect on the collected video fields. It is recommended for future research to establish whether faster and better performing hard drives and storage controllers affect the problem with lost video fields. The following are some recommended hardware testing configurations for computer hard drives:
1. Test SATA type I storage controller vs. SATA type II storage controller,
2. Test SATA type II storage controller vs. IDE ATA storage controller,
3. Test the impact of having different cache memory levels for hard drives – 2 Mb vs. 8 Mb, 8Mb vs. 16 Mb on the collection of video frames.